\documentclass[twocolumn, journal]{IEEEtran}

\IEEEoverridecommandlockouts
\usepackage[table]{xcolor}
\usepackage{color}
\usepackage{graphicx}
\usepackage{amsmath}
\usepackage{amssymb}
\usepackage{algorithm}
\usepackage{algorithmic}
\usepackage{amsmath}
\usepackage{multirow}
\usepackage{booktabs}
\usepackage{array}
\usepackage{amsthm}
\usepackage{stfloats}
\usepackage{caption}
\usepackage{subfigure}
\usepackage{bm}
\usepackage{setspace}
\usepackage{chngpage}
\usepackage{pifont}

\newcommand{\be}{\begin{equation}}
\newcommand{\ee}{\end{equation}}
\newcommand{\bea}{\begin{eqnarray}}
\newcommand{\eea}{\end{eqnarray}}
\newcommand{\ba}{\begin{array}}
\newcommand{\ea}{\end{array}}

\allowdisplaybreaks[4]
\title{Multi-Domain Optimization Framework for ISAC: From Electromagnetic Shaping to Network Cooperation
\thanks{R. Liu, M. Zafari, and A. L. Swindlehurst are with the center for Pervasive Communications and Computing, University of California, Irvine, CA 92697, USA (e-mail: rangl2@uci.edu; mzafarid@uci.edu; swindle@uci.edu).}
\thanks{M. Li is with the School of Information and Communication Engineering, Dalian University of Technology, Dalian 116024, China (e-mail: mli@dlut.edu.cn).}
\thanks{Bj\"{o}rn Ottersten is with the Interdisciplinary Center for Security, Reliability and Trust (SnT), University of Luxembourg, 1855 Luxembourg City, Luxembourg (e-mail: bjorn.ottersten@uni.lu)}
}
\author{Rang Liu,~\IEEEmembership{Member,~IEEE,}
        Ming Li,~\IEEEmembership{Senior Member,~IEEE,}
        Mehdi Zafari,~\IEEEmembership{Graduate Student Member,~IEEE,} \\
        Bj\"{o}rn Ottersten,~\IEEEmembership{Fellow,~IEEE,}
        and A. Lee Swindlehurst,~\IEEEmembership{Fellow,~IEEE}
        }
\begin{document}
\maketitle
\thispagestyle{empty}
\begin{abstract}
Integrated sensing and communication (ISAC) has emerged as a key feature for sixth-generation (6G) networks, providing an opportunity to meet the dual demands of communication and sensing. Existing ISAC research primarily focuses on baseband optimization at individual access points, with limited attention to the roles of electromagnetic (EM) shaping and network-wide coordination. The intricate interdependencies between these domains remain insufficiently explored, leaving their full potential for enhancing ISAC performance largely untapped. To bridge this gap, we consider multi-domain ISAC optimization integrating EM shaping, baseband processing, and network cooperation strategies that facilitate efficient resource management and system-level design. We analyze the fundamental trade-offs between these domains and offer insights into domain-specific and cross-domain strategies contributing to ISAC performance and efficiency. We then conduct a case study demonstrating the effectiveness of joint multi-domain optimization. Finally, we discuss key challenges and future research directions to connect theoretical advancements and practical ISAC deployments. This work paves the way for intelligent and scalable ISAC architectures, providing critical insights for their seamless integration into next-generation wireless networks.
\end{abstract}


\section{Introduction: The Need for Multi-Domain Optimization in ISAC}

Rapid expansion of wireless services and the ubiquitous proliferation of smart devices have intensified the demand for high-speed connectivity and precise environmental awareness. Recent 3GPP feasibility studies and emerging service requirements for Release 19 \cite{3GPP1}, \cite{3GPP2} emphasize integrated sensing and communication (ISAC) as a potential vertical for sixth-generation (6G) networks. ISAC can support diverse applications from enhancing traffic safety using automotive radar to high-accuracy localization in dense urban environments, all while maintaining reliable wireless connectivity. By enabling communication signals to simultaneously function as radar-like probes, ISAC can reduce hardware costs and alleviate spectrum congestion, unlocking transformative capabilities for advanced applications such as autonomous vehicular networks, industrial IoT, smart infrastructure, and real-time environmental monitoring. This dual functionality not only improves spectral and energy efficiency but also enables more robust and intelligent network operations, positioning ISAC as attractive feature for future wireless networks \cite{FLiu-JSAC-2022}.

Despite these benefits, ISAC systems face challenges due to inherent differences in target sensing and wireless communications that impact system and waveform designs. Communication generally requires moderate transmit power and stable data throughput, whereas sensing demands higher power to detect distant or weakly reflective targets. This asymmetry exacerbates self-interference, especially in monostatic configurations. Furthermore, communication systems favor short symbol durations for low latency and longer frames for error correction, while sensing requires longer coherent intervals for precise velocity estimation and short signal pulses for closely spaced echo resolution. Sensing also typically requires wider bandwidths to achieve high range resolution. These differences complicate resource allocation, hardware design, and interference management, necessitating comprehensive optimization approaches beyond conventional waveform-centric solutions.

The integration required by ISAC systems must not compromise communication performance, ideally enhancing it through location-based information. This requires the imposition of stringent constraints due to the required sharing of wireless resources. Common resource-sharing methods include time-division duplex (TDD) and frequency-division duplex (FDD) techniques, together with the partitioning of spatial resources. TDD simplifies hardware implementation but introduces challenges such as self-interference due to rapid transmit-receive switching. FDD naturally reduces interference through separate frequency bands but increases hardware complexity. Spatial resource partitioning allocates separate antennas or beams to different tasks to achieve high spectral efficiency, but suffers from implementation complexities. Practical ISAC systems typically combine these approaches to balance performance trade-offs.

\begin{figure*}[!t]
    \centering
    \includegraphics[width=\linewidth]{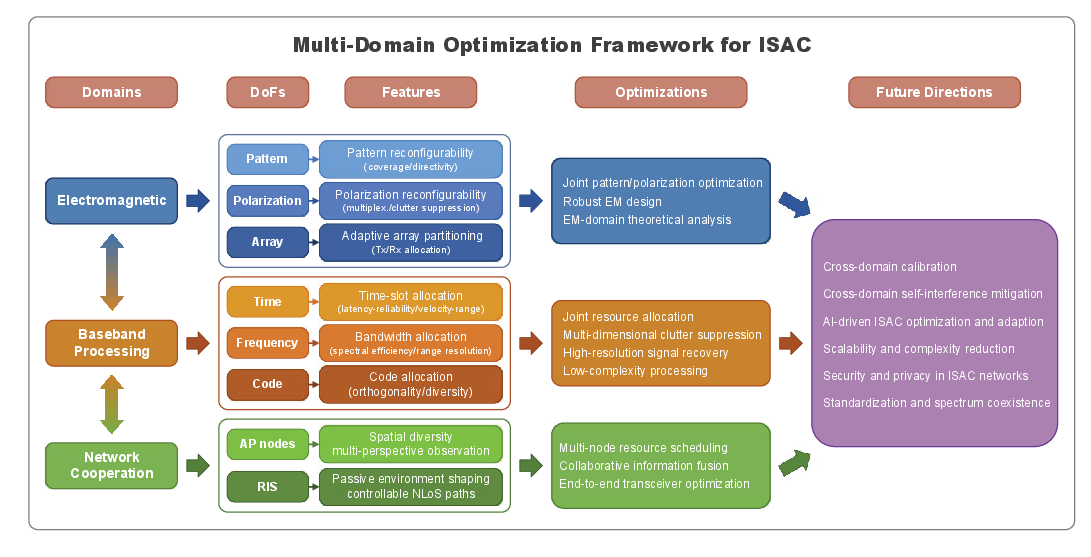}
    \vspace{-0.2 cm}
    \caption{Multi-domain optimization framework for ISAC.}
    \label{fig:outline}\vspace{-0.0 cm}
\end{figure*}
Although these resource-sharing strategies have been extensively studied, existing research has predominantly focused on waveform and beamforming optimization at the baseband and RF stages \cite{XLiu-TSP-2020}, refining signal characteristics such as amplitude, phase, frequency, and timing. While effective, these approaches do not fully consider the electromagnetic (EM) domain, where the antenna array configuration determines critical properties of the radiated signals and array response. Given the distinct propagation characteristics of radar targets and communication users, neglecting EM-level optimization constrains system flexibility. Therefore, dynamic antenna designs and EM-aware radiation pattern control are necessary to fully exploit ISAC capabilities.

In practical scenarios, ISAC systems encounter significant challenges including interference from clutter, weak target returns, and ambiguous feature extraction, particularly in complex environments such as low-altitude urban unmanned aerial vehicle (UAV) monitoring. In these scenarios, multipath reflections from buildings and vehicles generate dynamic clutter, while small UAVs with small radar cross sections (RCS) produce near-noise-floor echoes, limiting the performance of conventional detection and classification algorithms. 
Interference characteristics also strongly depend on the system configuration; monostatic setups inherently suffer self-interference due to co-located transmit and receive arrays, whereas bistatic or multistatic architectures spatially separate these functions, reducing interference at the expense of increased complexity. Effective interference cancellation, exploiting multiple dimensions including time, frequency, and space, and potentially employing advanced full-duplex technologies, is thus crucial for accurate sensing. Distributed or multistatic ISAC architectures mitigate self-interference through spatial separation, leverage spatial-geometric diversity, and enable coherent multi-perspective signal fusion. Such cooperative frameworks can improve clutter suppression and target localization accuracy \cite{KMeng-TWC-2025}, making them suitable for mission-critical applications such as aerial surveillance and industrial automation.

This paper proposes a systematic multi-domain optimization framework integrating EM shaping, baseband processing, and network cooperation. As illustrated in Fig.~\ref{fig:outline}, this framework enables cross-layer coordination, from physical components to network-level operations. Specifically, in the EM domain, flexible control of beam patterns and polarization enhances sensing accuracy and communication throughput. The baseband processing domain focuses on waveform and beamforming optimization, together with resource allocation across the time, frequency, and code dimensions. The network cooperation domain leverages multi-node architectures, resource scheduling, and network-aware processing to improve spatial diversity, adaptability, scalability, and cooperative performance in distributed ISAC systems. Sections II–V of the paper introduce relevant performance metrics and optimization strategies for these domains. Through dynamic coordination, the proposed framework enables more intelligent resource management, effective interference mitigation, and superior sensing–communication trade-offs, particularly in complex and rapidly evolving wireless environments. This is illustrated in Section V using a case study demonstrating the effectiveness of joint multi-domain optimization. Future research directions and conclusions are then given in Sections VI and VII.

\begin{figure*}[!t]
    \centering
    \includegraphics[width=0.93\linewidth]{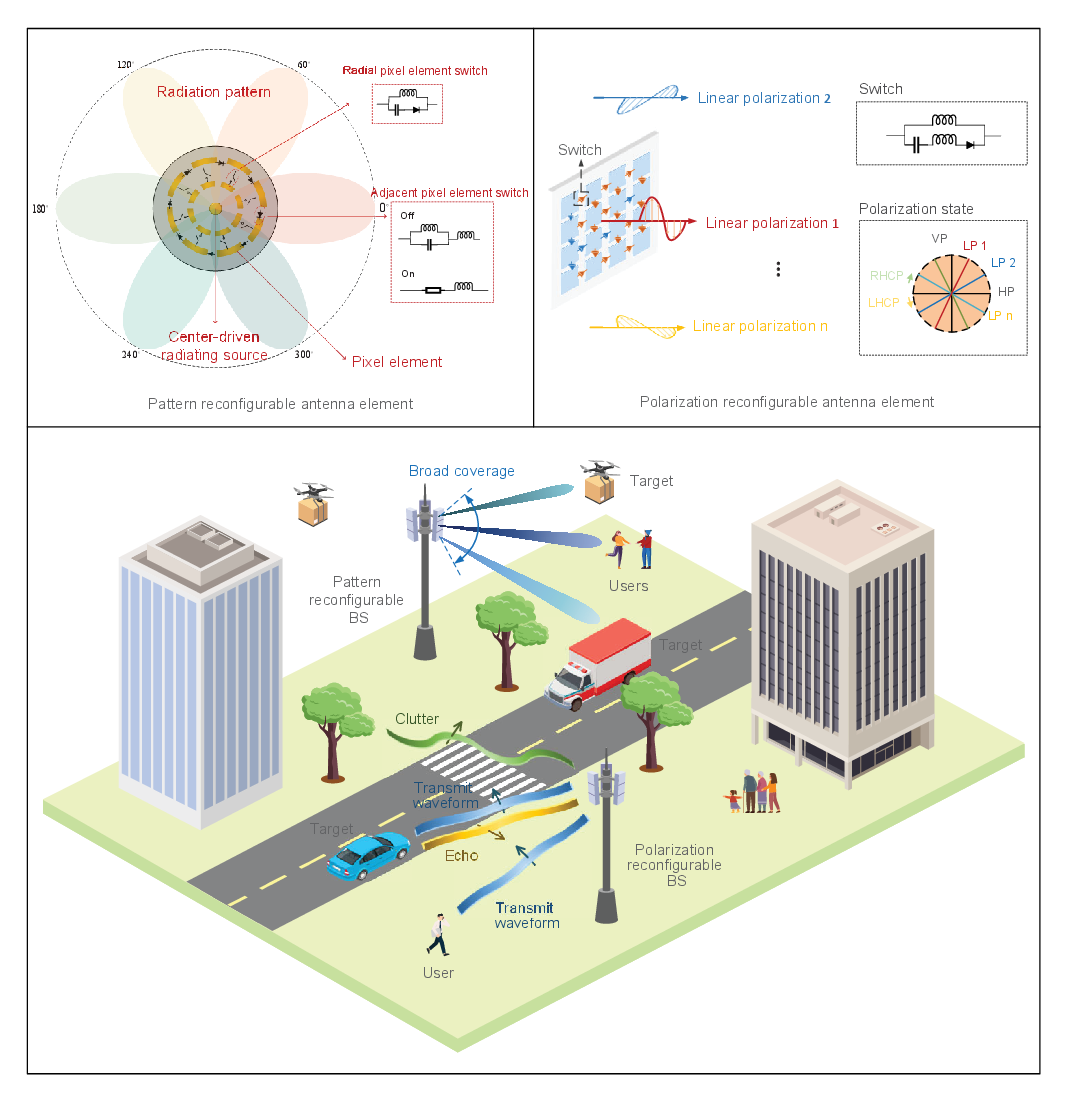}
    \vspace{-0.0 cm}
    \caption{Electromagnetic domain optimization.}
    \label{fig:EML}\vspace{0.0 cm}
\end{figure*}
\section{Electromagnetic Domain Optimization: Reconfigurable Antennas and EM Field Control}

\subsection{EM-Related Optimization and Reconfigurable Antennas}

ISAC systems require antenna arrays capable of simultaneously supporting high-precision sensing and high-throughput communication. However, challenges arise due to fundamental differences in the properties of radar targets and communication users. The location, orientation, and mobility of a communication user (``user equipment'', UE), is typically different from radar targets such as UAVs or vehicles. This disparity results in heterogeneous propagation channels characterized by distinct multipath profiles and angle/Doppler spreads. 
Conventional antennas with fixed patterns and polarizations struggle to adapt effectively to these heterogeneous conditions, limiting both sensing performance and spectral efficiency. Although multipath propagation can enhance communication via spatial diversity, environmental clutter reflections hinder target sensing by masking weak echoes. Consequently, optimizing EM-domain factors, including adaptive antenna radiation control, interference mitigation, and spatial resource allocation, is crucial.

Reconfigurable antennas address some of these challenges by introducing EM-domain adaptability through dynamic control of the antenna and array characteristics. Unlike traditional antennas with fixed radiation characteristics, reconfigurable or ``fluid'' antenna arrays integrate tunable elements such as electronically steerable parasitic arrays, frequency selective surfaces, and pixel-based structures, to enable dynamic adaptation of the antenna radiation patterns \cite{YZhang-TWC-2022}, frequency responses, and polarization states \cite{WZheng-TAP-2024}. By leveraging these capabilities, intelligent reconfigurable antennas can optimize spatial energy distribution, mitigate interference, and enhance spectral utilization.
Fig.~\ref{fig:EML} illustrates examples of reconfigurable antenna elements, highlighting their pattern and polarization reconfigurability capabilities alongside corresponding ISAC application scenarios.  

\subsection{Radiation Pattern Optimization for ISAC}
Antenna radiation patterns significantly impact ISAC performance, as they control the spatial distribution of transmit/sensed energy. Traditional antennas typically provide either omnidirectional coverage with limited directivity or directional patterns with restricted coverage. However, ISAC applications benefit from adaptive transmit radiation patterns due to the distinct spatial distributions of UEs and targets. 
Ground-based UEs require high-gain directional beams for high throughput, whereas coverage of airborne or mobile radar targets benefits from broader beams for reliable sensing. As depicted in Fig.~\ref{fig:EML}, pattern-reconfigurable antennas facilitate flexible energy distribution, enabling broad coverage for reliable detection and adaptive beamforming for reliable tracking of moving targets. Reconfigurable antennas address this trade-off by dynamically adjusting radiation characteristics to ensure efficient power distribution, balancing coverage and energy efficiency beyond the limitations of conventional phased arrays.

\subsection{Polarization Control for ISAC}
Polarization control introduces an additional degree of freedom (DoF) for ISAC. Polarization-division multiplexing (PDM) exploits orthogonal polarization states to support multiple independent data streams within the same frequency band, boosting spectral efficiency. However, traditional PDM implementations require dedicated transceivers for each polarization, increasing hardware complexity and cost. Polarization-reconfigurable antennas address this challenge using single-feed designs and tunable feeding networks, achieving polarization reconfigurability without significantly increasing the required RF hardware. As shown in Fig.~\ref{fig:EML}, polarization reconfigurability enables target echoes to be more easily distinguished from clutter based on polarization-dependent scattering characteristics, improving detection and classification performance. By jointly optimizing polarization control at both the transmitter and receiver, ISAC systems can effectively balance communication throughput and sensing accuracy while accommodating practical constraints such as limited RF chains and hardware complexity.

\subsection{Array Partitioning for ISAC}
Array partitioning provides additional DoFs for monostatic architectures, where transmit and receive functions share a common antenna array. Typically, the array is divided into two subarrays: one transmitting dual-functional signals for sensing and communication, and the other dedicated to receiving sensing echoes. Effective array partitioning influences beamforming gain, angular coverage, and sensing accuracy. Allocating more antennas for transmission enhances both communication and sensing performance, improving spectral efficiency, spatial multiplexing capability, and communication range, while also providing greater beamforming gain and more precise target illumination. Conversely, assigning more antennas for reception increases the effective aperture for sensing, improving spatial resolution and direction-of-arrival (DoA) estimation accuracy \cite{RLiu-TWC-2025}. However, suboptimal partitioning and antenna placement can lead to increased mutual coupling, grating lobes, and self-interference. Adaptive array partitioning dynamically configures the antenna assignment based on operational conditions, mitigating self-interference and optimizing the balance between communication throughput and sensing precision.

\subsection{Joint EM Field Processing and Optimization for Robust ISAC}

\textbf{Joint Optimization of Radiation Pattern and Polarization:}
Radiation pattern and polarization are fundamentally coupled EM antenna properties. Radiation pattern adjustments impact polarization purity and cross-polarization isolation, while polarization configurations influence spatial beamforming quality and directional gain. Jointly optimizing these parameters requires advanced algorithms that simultaneously manage complex trade-offs, such as balancing polarization multiplexing gains against cross-polarization interference levels, and ensuring wide-beam illumination for sensing without compromising polarization discrimination. By explicitly managing the inherent electromagnetic interdependencies between radiation patterns and polarization states, ISAC systems can achieve substantial improvements in both spectral efficiency and clutter suppression.

\textbf{Robust EM Design Under Hardware Impairments:}
Although reconfigurable antennas can improve EM-domain flexibility, practical ISAC performance is inherently constrained by hardware imperfections. Mutual coupling among densely packed antenna elements, radiation pattern and polarization state variability, and amplitude/phase mismatches in tunable components lead to distorted beams, increased sidelobe levels, and increased cross-polarization interference. Communication systems typically exhibit moderate sensitivity to such imperfections due to robust decoding mechanisms and statistical channel modeling. However, sensing performance is significantly more vulnerable, as accurate extraction of target parameters (e.g., angle, range, Doppler) critically depends on precise parameter-based channel estimation and accurate modeling of antenna array responses. Even minor hardware-induced deviations can amplify estimation errors, causing beam misalignment, diminished main-lobe gain, and elevated sidelobes. Moreover, limited isolation between transmit and receive paths exacerbates self-interference, reducing the ability to detect weak targets. Thus, rigorous array calibration, precise array response modeling, and hardware-aware optimization with real-time compensation are essential to maintaining robust sensing accuracy.

\textbf{Theoretical Performance Analysis in the EM Domain:}
Electromagnetic information theory \cite{JZhu-WC-2024} provides a framework for rigorously analyzing the impact of spatial and polarization DoFs, establishing capacity limits for communication and resolution bounds for sensing. Channel capacity analyses and Cramér-Rao bounds (CRBs) define theoretical performance limits, while analyses based on Maxwell's equations offer insights into antenna/wave interactions and propagation effects. These analytical tools can help establish fundamental performance bounds and provide practical insights.

 \begin{figure*}[!t]
    \centering
    \includegraphics[width=0.78\linewidth]{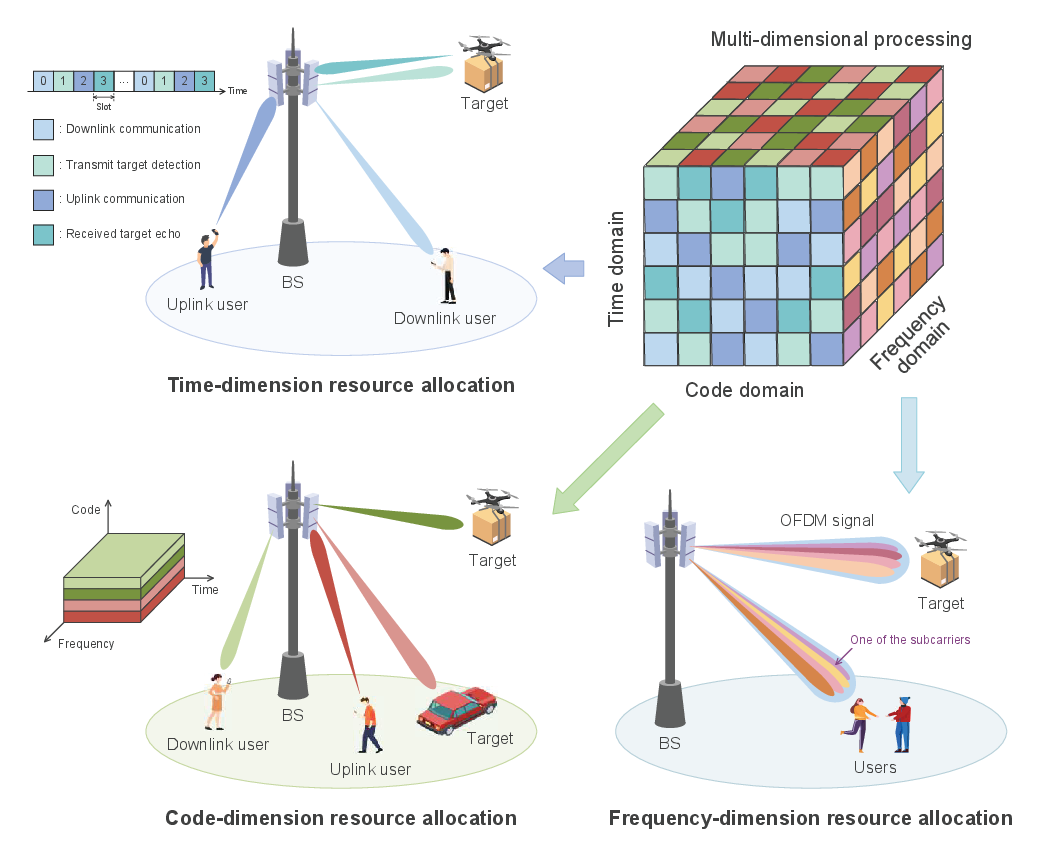}
    \vspace{-0.0 cm}
    \caption{Baseband processing domain optimization.}
    \label{fig:SPL}\vspace{-0.0 cm}
\end{figure*}

\section{Baseband Processing Domain Optimization: Efficient Resource Allocation and Joint Processing}

\subsection{Key Challenges for ISAC Baseband Processing}
The integration of sensing and communication functionalities introduces significant challenges at the baseband, primarily due to limited shared resources across the time, frequency, code, power, and space dimensions. Challenges arise due to interference between communication and sensing signals, clutter from environmental reflections, and multi-user interference, especially in dense networks. Effectively addressing these challenges requires careful multi-dimensional resource allocation at the transmitter (TX), coupled with sophisticated multi-dimensional signal processing techniques at the receiver (RX). As illustrated in Fig.~\ref{fig:SPL}, ISAC systems can flexibly allocate distinct time slots, orthogonal codes, and frequency carriers to communication and sensing tasks, resulting in multi-dimensional echo signals at the RX.

At the TX, conventional approaches prioritize beamforming and power allocation to optimize communication-centric metrics. However, time, frequency, and code—domain processing are equally important and remain relatively underexplored. Joint resource allocation across these dimensions can enhance spectral efficiency and reduce interference. At the RX, reliably extracting weak sensing echoes amid strong clutter and interference requires joint multidimensional signal processing across spatial, temporal, and spectral domains. However, practical RX architectures impose strict hardware constraints, including limited RF chains and computational complexity, restricting processing flexibility. Consequently, RX methods must jointly exploit these dimensions while carefully considering hardware and architectural limitations.

\subsection{Time-Domain ISAC Resource Allocation}
Time-domain resource allocation for ISAC typically involves assigning dedicated time slots for sensing and communication. In principle, this enables optimized waveform designs for each function \cite{PLi-2025}. However, practical communication system constraints such as standardized frame structures and latency requirements restrict temporal flexibility. In addition, sensing and communication operate at inherently conflicting timescales and rely on different types of waveforms. 
Communication systems transfer information in short duration packets to satisfy latency demands, whereas sensing requires integration over longer coherent intervals for accurate estimation of parameters that change relatively slowly. Furthermore, communication signals vary randomly in time, while radar signals follow deterministic patterns.
Thus, exploiting time-domain flexibility in ISAC systems requires careful management of timescale mismatches and waveform constraints.

\subsection{Frequency-Domain ISAC Resource Allocation}
Frequency-domain resource allocation is crucial for simultaneously meeting sensing resolution and accuracy requirements, and achieving high communication spectral efficiency. Allocating wider bandwidths to sensing improves range estimation accuracy but correspondingly reduces available spectral resources for communication. Fortunately, typical ISAC scenarios involve a limited number of targets, allowing compressed sensing and sparse frequency-domain processing for range estimation, thereby freeing additional bandwidth for communication. 
Practical frequency-domain allocation strategies must also consider interference arising from differing transmit powers: sensing signals typically employ high transmit power to detect distant or weak targets, whereas communication signals operate at relatively lower power levels to control adjacent cell interference. This power difference can lead to leakage from sensing into communication bands. Thus, effective frequency-domain allocation strategies must carefully balance sensing accuracy, spectral efficiency, and interference to fully leverage the frequency-domain flexibility of ISAC systems.

\subsection{Code-Domain ISAC Resource Allocation}
Exploiting DoFs available from waveform coding further improves communication reliability and sensing accuracy by effectively managing waveform orthogonality and interference. In communication, coding minimizes mutual interference, enabling efficient multi-user access, robust data transmission, and improved spectral efficiency. For sensing, coded waveforms enhances diversity \cite{JLi-SPM-2007}, reducing autocorrelation sidelobes and improving range and Doppler resolution. Proper code assignment across time-frequency resource blocks enables effective target separation without compromising communication reliability. However, imperfect code orthogonality and channel variations degrade practical ISAC performance. Adaptive code allocation dynamically optimizes codes according to real-time channel conditions and sensing demands, improving estimation accuracy and resource utilization while mitigating cross-correlation interference.

\subsection{Joint Multi-Dimensional Processing in ISAC}
While individual optimizations in the time, frequency, or code dimensions enhance specific ISAC capabilities, effectively managing their inherent trade-offs requires joint resource allocation at the TX and integrated multi-dimensional processing at the RX. At the TX, joint optimization across all dimensions is essential for interference mitigation and efficient resource utilization. At the RX, practical hardware constraints necessitate methods that jointly exploit the spatial, temporal, and spectral domains for robust clutter suppression, high-resolution signal recovery, and low-complexity signal extraction, ensuring reliable sensing of weak echoes.

\textbf{Multi-Dimensional Adaptive Clutter Suppression:} Interference and clutter often exhibit strong correlations across multiple dimensions, limiting the effectiveness of one-dimensional suppression techniques \cite{RLiu-JSAC-2022}. Strong clutter sources that mask weak targets span multiple range-angle bins, complicating detection and parameter estimation. Multi-dimensional processing leverages space-time-frequency-polarization diversity to jointly suppress clutter while preserving weak target signals. In particular, Doppler-based filtering differentiates stationary clutter from moving targets, while polarization-based techniques leverage differences between target and clutter scattering characteristics. Integrating these techniques with adaptive filtering enhances target detection and parameter estimation.

\textbf{High-Resolution Signal Recovery for ISAC:} Sensing waveforms in communication-centric ISAC systems are inherently limited by narrow communication bandwidths, standardized frame structures, and fixed sampling rates optimized primarily for communication, restricting resolution for closely spaced targets. To address this issue, high-resolution signal recovery techniques are essential. Super-resolution algorithms enable precise range and Doppler estimation beyond classical limits, while structured reconstruction methods including compressed sensing and deep-learning-based denoising effectively recover detailed target information from sub-Nyquist or coarsely sampled signals. Leveraging these advanced techniques facilitates adequate sensing performance without substantial changes to existing communication infrastructure.

\textbf{Low-Complexity Processing for ISAC:} Multi-dimensional joint optimization across space, time, and frequency requires processing large-scale data sets, increasing computational complexity and posing challenges for real-time implementations. To address this challenge, low-complexity baseband processing techniques exploit intrinsic signal characteristics such as sparsity \cite{ZXiao-Tcom-2025} and low-rank structures. 
Hybrid beamforming employs analog networks to decrease the number of RF chains and simplify baseband processing. However, these techniques introduce approximation errors that potentially degrade sensing accuracy. Practical ISAC systems thus benefit from combining computationally efficient processing with structured high-resolution reconstruction methods.

\begin{figure*}[!t]
    \centering
    \includegraphics[width=0.85\linewidth]{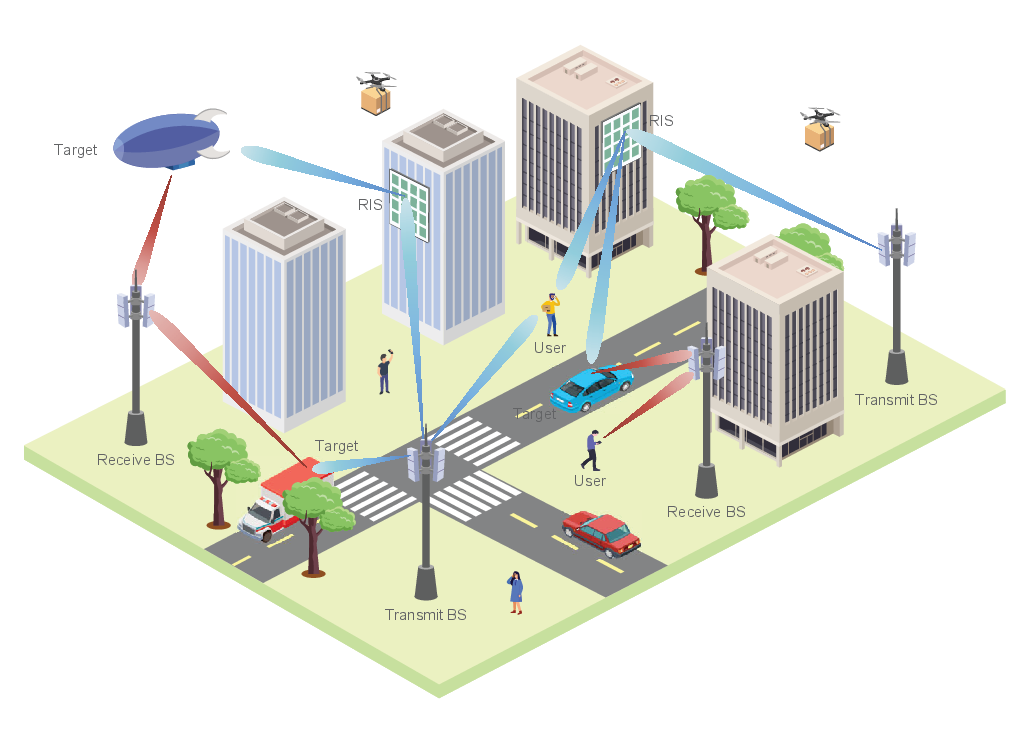}
    \vspace{-0.2 cm}
    \caption{Cooperative optimization at the network level.}
    \label{fig:NtwL}\vspace{-0.0 cm}
\end{figure*}

\section{Network Cooperation Domain: \\ Multi-Node Configuration and Optimization}

\subsection{Multi-Node Cooperation for ISAC}
Single-node monostatic ISAC systems are inherently constrained by limited spatial coverage, angular visibility, and severe self-interference, degrading both communication reliability and sensing performance, especially for low-RCS targets. Multi-node cooperative architectures address these limitations by leveraging spatially distributed nodes, achieving extended coverage, enhanced multi-perspective sensing, and reduced interference sensitivity. Fig.~\ref{fig:NtwL} illustrates some potential cooperative ISAC networks in which distributed access points (APs) and reconfigurable intelligent surfaces (RIS) provide diverse observation angles, alleviating signal blockages and improving target detection and localization accuracy.
Despite their advantages, effective multi-node cooperation introduces challenges such as optimal node configuration, resource scheduling across multiple nodes, efficient fusion of collaborative information, and end-to-end joint optimization.

\subsection{AP Node Configuration in ISAC Networks}
Flexible AP node configuration provides additional spatial DoFs that can enhance ISAC performance through increased spatial diversity. The ability to strategically assign the nodes as transmitters or receivers provides ISAC systems with distinct target observation perspectives that substantially improve sensing robustness, particularly since target parameter estimation accuracy depends strongly on the observation geometry. Similarly, spatially distributed AP configurations alleviate signal blockages, multipath fading, and interference effects, enhancing communication link reliability and spatial multiplexing capabilities. An effective AP configuration thus involves dynamically balancing sensing and communication objectives, considering environmental effects, mobility patterns, and energy constraints. By fully exploiting this spatial flexibility, optimal AP node assignment significantly improves ISAC system coverage, sensing accuracy, and communication robustness in dynamic scenarios.

\subsection{RIS Configuration in ISAC Networks}
RIS nodes enrich the number of spatial DoFs in ISAC networks by  passively reshaping wireless environments. By dynamically creating controllable NLoS paths, an RIS can extend coverage, reduce blind zones, and enhance spatial diversity, improving sensing robustness and communication reliability. Unlike active AP nodes, an RIS operates with minimal low-power RF hardware, offering scalability and energy-efficiency. However, integrating RIS into ISAC networks introduces unique challenges such as optimal RIS-to-AP association, real-time adaptive reflection optimization to enhance angular resolution and mitigate interference, and efficient network-wide coordination with minimal computational overhead. Developing adaptive, low-complexity RIS control algorithms that balance sensing and communication requirements is thus essential.

\subsection{Multi-Node Cooperative Configuration and End-to-End ISAC Optimization}

\textbf{Multi-Node Resource Scheduling:} Multi-node resource scheduling dynamically assigns transmission, reception, and reflection roles across AP and RIS nodes, affecting sensing geometry and communication link quality. Optimal role assignment must flexibly adapt to real-time environmental dynamics, mobility patterns, and interference conditions. Scheduling complexity also varies by duplexing or resource-sharing methods. TDD-based ISAC systems require precise synchronization and careful time-slot allocation, since even minor synchronization errors during coherent multi-node sensing can degrade target parameter estimation accuracy. Similarly, FDD-based systems necessitate coordinated frequency-band allocation across nodes to efficiently use the available spectrum, while spatial resource partitioning can reduce self-interference due to limited transmit–receive isolation. Therefore, resource-sharing and scheduling strategies must comprehensively consider practical communication and sensing requirements, channel conditions, interference characteristics, as well as node-specific hardware and architectural constraints.

Another critical challenge is handover management, as ISAC requires simultaneously preserving communication continuity and sensing consistency. Inefficient handover strategies can disrupt target tracking, degrade sensing accuracy, and introduce latency or spectral inefficiencies. Effective multi-node scheduling strategies should carefully coordinate node-role transitions, RIS configurations, and collaborative fusion.

\textbf{Collaborative Sensing and Information Fusion:} Multi-node ISAC networks leverage geometric diversity by coherently fusing multi-perspective UE and target observations to improve target detection and localization.   However, practical challenges arise from heterogeneous node configurations, including hardware mismatches (e.g., varying antenna gains, phase noise characteristics), asynchronous sampling rates, and non-uniform channel conditions. These factors lead to spatio-temporal misalignment and calibration errors, potentially degrading sensing performance unless addressed through robust cross-node fusion strategies. 
Effective information fusion must handle node differences, redundancy, and latency constraints.

Information fusion typically occurs at three distinct levels. Signal-level fusion centrally aggregates raw in-phase/quadrature signals, enabling coherent processing and high-precision localization, but demands stringent synchronization and substantial fronthaul capacity. Feature-level fusion exchanges compressed parameters (e.g., DoA, delay, Doppler shifts), balancing communication overhead with accuracy. Decision-level fusion integrates local detection outcomes, further minimizing overhead and improving robustness to node-level variations.

Distributed computational frameworks such as message passing, belief propagation, and consensus-based estimation, enable decentralized, iterative refinement of sensing information, ensuring consistency despite local variations. Additionally, hierarchical fusion can reduce latency and computation by preprocessing measurements at edge nodes before centralized fusion. Leveraging these adaptive strategies can help achieve robust, scalable, and efficient multi-node ISAC performance.

\textbf{End-to-End Joint Optimization of ISAC Transceiver Architectures}: Multi-node resource scheduling optimizes transmission-side operations, while collaborative sensing enhances reception-side information fusion. Treating these processes independently, however, can cause misaligned objectives and suboptimal overall performance. An end-to-end joint optimization framework addresses this issue by integrating transceiver selection, waveform design, resource scheduling, and information fusion. This unified approach enables adaptive sensing–communication trade-offs, mitigates interference, and enhances both spectral efficiency and sensing accuracy.

Given the complexity of fully joint transceiver optimization, iterative refinement approaches provide practical alternatives. Receivers can extract critical sensing parameters and share these with transmitters, enabling dynamic waveform, beamforming, and power adjustments. 
Alternatively, learning-based methods such as graph neural networks (GNNs), graph attention networks (GATs), and reinforcement learning (RL), can approximate solutions obtained through joint optimization by exploiting network topology and real-time feedback, further improving adaptability and robustness.

\begin{figure}
    \centering 
    \includegraphics[width=\linewidth]{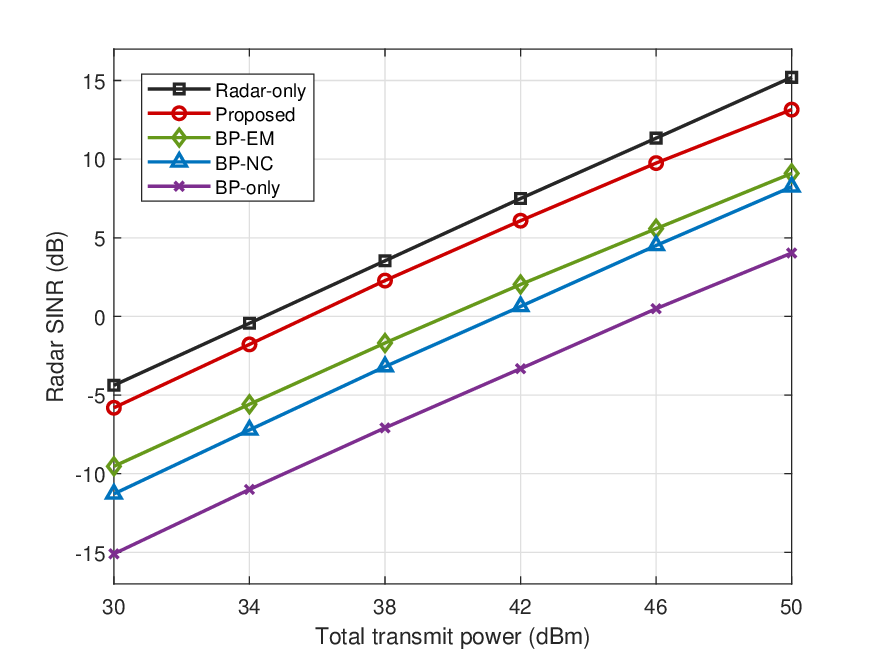}
    \vspace{-0.2 cm}
    \caption{Radar SINR vs.~total transmit power. ``Proposed'': Proposed multi-domain optimization of AP selection, subcarrier allocation, beamforming, and polarization across network, baseband, and EM domains. ``Radar-only'': Employs multi-domain optimization without communication constraints. ``BP-only'': Baseband-only optimization of subcarrier allocation and beamforming with fixed AP selection and polarization. ``BP-EM'': Joint optimization of subcarrier allocation, beamforming, and polarization with fixed AP selection. ``BP-NC'': Joint optimization of subcarrier allocation, beamforming, and AP selection with fixed polarization.)}
    \label{fig:simulation}\vspace{-0.0 cm}
\end{figure}

\section{Multi-Domain Optimization: A Case Study}

To demonstrate the effectiveness of the proposed multi-domain ISAC optimization framework, we study a cell-free MIMO-OFDM system employing polarization-reconfigurable antenna arrays. We consider a scenario with $6$ distributed APs cooperatively serving $8$ single-antenna UEs while detecting a single radar target. Each AP is equipped with a uniform linear array consisting of $4$ single-port polarization-reconfigurable antennas. The APs share a common OFDM frame structure composed of $128$ subcarriers, each having a frequency spacing of $120$ kHz. Each AP can function either as a transmitter (TX-AP) or receiver (RX-AP). Each TX-AP dedicates a tunable subset of subcarriers exclusively to radar sensing, whereas the remaining subcarriers are collectively employed by all TX-APs for communication. The optimization goal is to maximize radar SINR while satisfying constraints on communication sum-rate and transmit power. 
To solve this complicated problem, we decompose it into AP/subcarrier resource allocation and beamforming/polarization optimization subproblems and develop efficient algorithms to solve them.  
As shown in Fig.~\ref{fig:simulation}, the proposed multi-domain optimization framework significantly outperforms benchmarks that optimize over only a subset of the possible DoFs. The proposed approach is able to meet the downlink communication requirements with only a slight penalty in radar sensing performance. These results confirm the advantages of cross-domain optimization across EM, baseband, and network domains.

\section{Future Directions}

Fully unlocking the potential of ISAC in practical deployments requires addressing fundamental cross-domain optimization challenges. Decoupled designs that separately optimize electromagnetic shaping, baseband processing, or network cooperation, ignore their inherent interdependencies and trade-offs. Real-world ISAC scenarios lead to highly coupled optimization objectives, including conflicting sensing and communication performance criteria, nonlinear inter-domain interactions, scalability bottlenecks, and hardware constraints. Consequently, future research must embrace holistic, cross-domain methodologies that systematically integrate the electromagnetic, RF, baseband, and network domains, to ensure efficient, robust, and adaptive ISAC operation. 

\textbf{Cross-Domain Calibration and Hardware Imperfections:}
Calibration issues present critical cross-domain challenges that significantly limit practical ISAC sensing performance. As highlighted in Section II, antenna-array modeling/calibration errors such as mutual coupling and amplitude/phase mismatches distort beam patterns and degrade DoA estimation accuracy. Imperfect RF-chain calibration further introduces amplitude and phase distortion, reducing spatial resolution and sensing sensitivity. 
Timing synchronization errors compromise range and Doppler estimation, leading to reduced target resolution and increased false alarms. 
Inaccurate baseband channel estimation based on overly simplified models amplify these hardware-induced errors, diminishing parameter estimation quality. Future work must effectively address these challenges using integrated hardware-algorithm co-design for advanced calibration and compensation of hardware non-idealities.

\textbf{Cross-Domain Self-Interference Mitigation:}
Self-interference (SI), the leakage of transmit signals into sensitive receiver front-ends, poses a critical cross-domain challenge for practical ISAC deployments. Limited antenna isolation in monostatic or spatially partitioned arrays, duplexer imperfections causing frequency leakage in FDD systems, and transient leakage during TDD transmit-receive switching reduce receiver sensitivity and dynamic range. At the baseband level, SI issues are exacerbated by imperfect waveform designs, incomplete orthogonality of spreading codes, near-far effects, and inter-node interference in multi-node networks. These effects compromise sensing accuracy and degrade communication performance. Effective SI mitigation necessitates integrated cross-domain solutions combining enhanced antenna isolation techniques, advanced RF duplexing filters, adaptive baseband SI cancellation algorithms, and dynamic multi-node resource scheduling strategies.

\textbf{AI-Driven ISAC Optimization and Adaption:}
Traditional ISAC optimization relies heavily on model-based approaches, and may become computationally intractable in dynamic environments. Learning-based methods, including RL, GNNs, and self-supervised learning, offer promising alternatives for real-time adaptation, intelligent beamforming, and distributed resource management. Future research should explore hybrid model- and data-driven techniques for self-optimizing ISAC architectures capable of autonomous decision-making under uncertain and time-varying conditions.

\textbf{Scalability and Complexity Reduction:} Expanding ISAC to massive multi-input multi-output (MIMO), RIS-assisted, and cell-free architectures introduces significant computational challenges in beamforming, resource allocation, and interference management. Existing optimization frameworks often assume fixed system parameters, leading to global re-optimization when network conditions change. Such approaches are computationally prohibitive in large-scale, dynamic networks. Future ISAC architectures must support scalable and adaptive optimization, where network expansions and modifications can be handled with localized updates, reducing computational complexity and signaling overhead while maintaining system performance.

\textbf{Security and Privacy in ISAC Networks:} The integration of sensing and communication introduces new security and privacy risks, including eavesdropping via radar sensing, adversarial spoofing, and data leakage from ISAC signals \cite{ZRen-JSAC-2024}. Unlike traditional wireless security, ISAC requires dual-layer protection to ensure both secure data transmission and robust sensing. Future research should explore secure waveform design, privacy-preserving signal processing, and adversarial attack mitigation techniques to protect ISAC networks against emerging threats.

\textbf{Standardization and Spectrum Coexistence:} The deployment of ISAC in 6G networks requires addressing standardization, spectrum coexistence, and regulatory compliance. Future research should focus on ISAC-friendly spectrum-sharing mechanisms, protocol standardization, and coexistence strategies with legacy communication systems. Ensuring interoperability across heterogeneous networks is essential for the adoption of ISAC in applications such as autonomous systems, smart infrastructure, and industrial IoT.

\section{Conclusions}
This article has introduced the concept of multi-domain optimization for improving ISAC system performance through cross-domain coordination. By bridging the gaps between EM resource management, baseband processing, and network cooperation, we have provided guidelines for research on optimizing ISAC systems across different domains. Looking ahead, it will be crucial to address scalability, security, and adaptability to enable deployment of working ISAC systems. Future advancements in intelligent optimization and standardization will be key to enabling efficient and scalable ISAC systems, paving the way for their seamless integration into next-generation wireless networks.

% \section{Acknowledgment}

% \centerline{\textbf{BIOGRAPHIES}}

\end{document}